\pdfoutput=1

\RequirePackage{fix-cm}
\RequirePackage{rotating}

\documentclass[smallextended]{svjour3}
\smartqed 
\usepackage{cite}
\usepackage{mathptmx}
\usepackage{amsmath,amssymb,amsfonts}
\usepackage{graphicx}
\usepackage{subcaption}
\usepackage{textcomp}
\usepackage{xcolor}
\usepackage{listings}
\usepackage{multirow}
\usepackage{wrapfig}
\usepackage{xcolor,colortbl}
\usepackage[flushleft]{threeparttable}
\usepackage{amsmath}
\usepackage{relsize}
\usepackage{pifont}
\usepackage[sort&compress,numbers]{natbib}
\usepackage[skins]{tcolorbox}
\usepackage{wasysym}
\usepackage{pgfplots}
\usepackage{changepage} 
\usepackage{enumerate}
\usetikzlibrary{shadows}
\usepackage{enumitem}
\usepackage[ruled,vlined]{algorithm2e}
\usepackage{algorithmicx}
\usepackage[hidelinks]{hyperref}
\usepackage{tcolorbox}
\newtcolorbox{blockquote}{colback=gray!5,boxrule=0.4pt,colframe=black,fonttitle=\bfseries}
\usepackage[T1]{fontenc}

\newcommand{\BLUE}{\color{black}}
\newcommand{\BLACK}{\color{black}}

\usepackage{cite}

\newcommand{\bi}{\begin{itemize}}
\newcommand{\ei}{\end{itemize}}
\newcommand{\be}{\begin{enumerate}}
\newcommand{\ee}{\end{enumerate}}

\usepackage{subcaption,siunitx,booktabs}

\pgfplotsset{compat=1.10}
\usepgfplotslibrary{fillbetween}

\def\BibTeX{{\rm B\kern-.05em{\sc i\kern-.025em b}\kern-.08em
    T\kern-.1667em\lower.7ex\hbox{E}\kern-.125emX}}
    
\definecolor{MyDarkBlue}{rgb}{0,0.08,0.45} 
\newcommand\MyBox[2]{
  \fbox{\lower0.75cm
    \vbox to 1.7cm{\vfil
      \hbox to 1.7cm{\hfil\parbox{1.4cm}{#1\\#2}\hfil}
      \vfil}%
  }%
}

\newcommand{\IT}{\sffamily{VEER}}

\begin{document}

\title{
VEER: Enhancing the Interpretability of Model-based
Optimizations%\\({\small in Multi-objective Configuration Optimization}
}

\author{Kewen Peng         \and
        Christian Kaltenecker \and
        Norbert Siegmund \and
        Sven Apel  \and
        Tim~Menzies
}

\institute{Kewen Peng,  Tim Menzies\\
Department of Computer Science, North Carolina State University, Raleigh, USA. \email{kpeng@ncsu.edu, timm@ieee.org}\\
\newline
Christian  Kaltenecker, Sven Apel\\
Department of Computer Science, Saarland University \& Saarland Informatics Campus, Germany. \email{kaltenec@cs.uni-saarland.de, apel@cs.uni-saarland.de}\\
\newline
Nobert Siegmund\\ 
Department of Computer Science, Leipzig University, Germany. Email: norbert.siegmund@informatik.uni-leipzig.de
}

\date{Received: date / Accepted: date}

\maketitle
\begin{abstract}
\small
{\em Background:} Many software systems can be tuned for multiple objectives (e.g., faster runtime, less required memory, less network traffic or  energy consumption, etc.). Such systems can   suffer from ``disagreement'' where different models  have different (or even opposite) insights and tactics on how to optimize a system. 
For configuration problems, we show that
(a)~model disagreement is rampant;
yet (b)~prior to this paper, it has barely been explored.

% This  paper shows that model disagreement can  be mitigated 
% via VEER, a one-dimensional approximation to the N-objective space. 
% Since it is exploring a simpler goal space,
% VEER runs very fast  (for eleven configuration problems).
% Even for our largest problem (with tens of thousands of possible configurations), VEER finds as good or better optimizations  with zero model disagreements, three orders of magnitude faster (since its one-dimensional output no longer needs the sorting procedure).  
% Based on the above, we recommend VEER as a very fast method to solve complex configuration problems, while at the same time avoiding   model disagreement. 

{\em Goal:} We aim at helping practitioners and researchers better solve  multi-objective configuration optimization problems, by resolving model disagreement.

{\em Method:} We propose a dimension reduction method called {\IT} that builds a useful one-dimensional approximation to the original N-objective space. Traditional model-based optimizers use Pareto search to locate Pareto-optimal solutions to a multi-objective problem, which is computationally heavy on large-scale systems. {\IT} builds a surrogate that can replace the Pareto sorting step after deployment.  

{\em Results:}  Compared to the prior state-of-the-art,
for 11 configurable systems, {\IT}   significantly 
reduces disagreement {\em and} execution time,
{\em without} compromising the optimization
performance in most cases. 
For our largest problem (with tens of thousands of possible configurations), optimizing with {\IT} finds as good or better optimizations with zero model disagreements, three orders of magnitude faster.

{\em Conclusion:} When employing model-based optimizers for multi-objective optimization, we recommend to apply {\IT}, which not only improves the execution time, but also resolves the potential model disagreement problem.


\section*{Conflict of interest}
We assert that the authors have no conflict of interest.

\end{abstract}
% \end{document}
 \newpage

\section{Introduction}\label{intro}

One of the recent successes of AI for software engineering (SE) is automated configuration~\cite{kaltenecker2020interplay}. Software comes with many options as well as various objectives, and exploring all these configuration options for multiple objectives can be tedious, time consuming, and even error-prone (when done manually). Much recent work shows that AI tools can dramatically improve this procedure; for instance, regression tree learners can report what subset of the configuration options are most influential to achieving better performance~\cite{golovin2017google,chen18}. Further, while it takes a long time to run the system with all possible (feasible) configuration settings, an incremental AI tool can reflect on the model learned so far to recommend what is the next most informative configuration to try~\cite{nair2018finding}. This way, the time required to run enough configurations to effectively optimize software can be substantially  reduced (e.g., as shown by the experiments of this paper, after running less than 100 configurations, we can optimize systems with nearly 90,000 configurations).

When business users ask   ``what has been learned from these models?'', we need {\em interpretable models} to offer comprehensible advice on how to best
configure a system~\cite{Sawyer11,tan2016defining,chen2018applications}. But when software systems have multiple objectives (e.g., faster transaction response time, fewer memory requirements, decreased network traffic, decreased energy consumption, etc), such advice could clash.   We call this the  {\em model disagreement problem}: while one model
shows some configuration options to be useful to achieve one objective, another model might argue that  such options are actually detrimental to another objective. Table~\ref{tab:example1}
shows examples of model disagreement. 
Later in this paper,
we show that model disagreement
is rampant in all our test cases (see our experimental
results in \S\ref{result}).

\begin{table}[!b]
\centering
% \scriptsize
\small
% \begin{tabular}{|cc|lc|}
% \hline
% \multicolumn{2}{|c|}{Model 1: maximize throughput} &
%   \multicolumn{2}{c|}{Model 2: reduce latency}\\ \hline
% \multicolumn{1}{|l|}{\multirow{4}{*}{Node 1}} &
%   spout $>$ 0.5 &
%   \multicolumn{1}{l|}{Node 1} &
%   sorters $>$ 0.09 \\ 
%   \cline{3-4}
% \multicolumn{1}{|l|}{} &
%   \cellcolor[HTML]{FFC300}{chunk $<$ 0.14} &
%   \multicolumn{1}{l|}{\multirow{3}{*}{Node 2}} &
%   sorters $<$ 0.09 \\
% \multicolumn{1}{|l|}{} &
%   \cellcolor[HTML]{FFC300}{max\_spout $>$ 0.55}  &
%   \multicolumn{1}{l|}{} &
%   \cellcolor[HTML]{FFC300}{max\_spout $<$ 0.05}\\
% %   \cline{1-2} \cline{5-6} 
% \multicolumn{1}{|l|}{} & \multicolumn{1}{l|}{} &
%   \multicolumn{1}{l|}{} &
%   \cellcolor[HTML]{FFC300}{chunk $>$ 0.05} \\ \hline
% \end{tabular}
\caption{Examples of model disagreement found by
 tools described later in this paper.
Please note that, in project SS-F, we mark the two rules $\mathit{chunck} < 0.14$ and $\mathit{chunck} > 0.05$ as disagreement because, although their ranges are not necessarily exclusive, they do point to the opposite optimizing directions (increase vs. decrease).}
\begin{tabular}{ccc}
% \hline
\toprule
Project & \multicolumn{1}{c}{Model 1}          & Model 2                                               \\ 
% \hline
\midrule
SS-E    & \multicolumn{1}{c}{How to minimize runtime} & How to minmize CPU load     \\ 
% \hline
% \midrule
        & columnTiling $= True$         & columnTiling $= False$                            \\
        & goodQuality $=  True$          & \multicolumn{1}{c}{goodQuality $= False$}        \\
        & AutoAltRef $= False$          & AutoAltRef $= True$                          \\ 
        % \hline
\midrule
SS-F    & \multicolumn{1}{c}{How to minimize latency} & How to maximize throughput \\ 
% \hline
% \midrule
        & max\_spout $>$ 0.55          & max\_spout $<$ 0.05                             \\
        & chunk $<$ 0.14                  & chunk $>$ 0.05                               \\ 
% \hline
\midrule
SS-G    & \multicolumn{1}{c}{How to minimize runtime} & How to minimize CPU load \\
% \hline
% \midrule
        & compressionZqap $= False$         & \multicolumn{1}{c}{compressionZqap $= True$} \\
        & compressionLzma $= True$     & \multicolumn{1}{c}{compressionLzma $= $ False}    \\
        & processorCount $>$ 0.17      & \multicolumn{1}{c}{processorCount $<$ 0.17}    \\ 
% \hline
\bottomrule
\end{tabular}
\label{tab:example1}
\end{table}

Looking at the literature, we find very little
  discussion on   model disagreement. That is,   model disagreement may be a   long-standing, but previously under-explored, problem.
This begs the questions ``why has this problem not been reported before?''  Perhaps
other researchers were content to stop after generating multiple solutions (e.g., 10,000 solutions across the   frontier of best solutions).
In our work, however, we have been studying interpretation tools that offer 
clear advice on how to best
configure a system. Hence, we prefer not to confuse users with a long list of candidate solutions. Instead, we prefer  rule-based summaries (such as those seen in Table~\ref{tab:example1}).

This paper tests the following conjecture.
If multiple objectives complicate interpretations, then one possible solution is to:
\begin{quote}{\em
(1)~Reduce  multi-dimensional objective space to a lower-dimensional space.  (2)~Reason in that reduced space.}
\end{quote}
We propose a tactic called {\IT}, which is applied as follows:  (a) lay out all the configuration options as points in an N-dimensional objective space, then (b) rank the best point as “1”. After that, we ``VEER'' to the nearest best point P to rank it as “2”;  and so on. The ranks found by VEERing across all the objectives are then used as a single-dimensional objective space. Configuration recommendations
are then found by reasoning over this  simpler space.

Overall, the contributions of this paper are:
\begin{itemize}[topsep=2pt]
    \item
    We verify the existence of the model disagreement problem in multi-objective software configuration.
    \item We show that model disagreement is 
{\bf not} a simple problem that can be easily solved via some  simplistic weighting mechanisms  (e.g.,  a naive multi-regression approach).
    \item 
    We propose a novel tactic to resolve the model disagreement problem and generate confusion-free model interpretations.
    \item 
    % Besides offering unequivocal interpretations, our tactic also
    We show that, since {\IT} is exploring
    a simpler goal space, it runs very fast  (up to 1,000 times faster while at the same time
    recommending configuration solutions that are as good as or better than the prior state-of-the-art).
\end{itemize}

% The rest of this paper is structured as follows. \S\ref{interpretability} introduced some fundamental ideas in interpretable AI, as well as insights we have seen from this field. \S\ref{background} discusses background knowledge of configurable software systems and multi-objective optimization. 
% % This section introduces some prior works in the related fields, including algorithms that are used and improved by our approach. 
% \S\ref{veer} describes the motivation, intuition, and algorithmic design of our approach, VEER. \S\ref{evaluation} presents research questions asked in this paper, along with evaluation process used to assess our answers to those questions. \S\ref{experiment} introduces data and learners used in our case studies, as well as performance criteria used to evaluate the experiment result.
% \S\ref{result} and \S\ref{discussion} report and discuss the experiment result respectively. The reliability and credibility of our results is discussed in \S\ref{threat}. Future work and directions are illustrated in \S\ref{futurework}. Finally, we conclude this work in \S\ref{conclusion}. 

\section{Background and Related Work}\label{rw}
\subsection{Why do configuration optimizations need    
``interpretation?''}

This section argues the necessity of interpretability and transparency for configuration models, which motivates this paper.

Configurable software systems come with numerous options such that users can customize the system for their varying requirements.
However, once a configuration space becomes large,
users can easily get confused by possible interactions among configuration options with distinct impacts on diverse objectives. 
For example, many software systems have poorly chosen defaults~\cite{van2017automatic,herodotou2011starfish}. Hence, it is useful to seek better configurations.
Unfortunately, understanding the configuration space of software systems with large configuration spaces can be challenging~\cite{xu2015hey,kolesnikov2019tradeoffs}, and
exploring more than just a handful of configurations is usually infeasible due to long benchmarking time~\cite{zhu2017optimized}.

When manual methods fail, automatic tools
can be of assistance. In the case of configuration,
those automatic tools are usually assessed with respect to
the objectives of the system. Many prior works have demonstrated effectiveness in optimizing configurations for single-objective and multi-objective systems~\cite{SiegmundGAK15,GuoYSASVCWY18, song2015multi}.  In this paper, while we focus on systems with more than one objectives, we add
one more assessment criteria: {\em transparency}.
A recent  2021 report by the Gartners group\footnote{ 
\url{http://tiny.cc/gartners21}} states
\begin{quote}
{\em 
Responsible AI governance (and)
{\bf transparency} (our emphasis) ... is the most valuable differentiator in this market, and every listed vendor is making progress in these areas.}
\end{quote}
A model with conflicting 
interpretations on different objectives is considered more of {\em opaque}
than {\em transparent}.
That is, model disagreement complicates transparency since users cannot directly and immediately understand the implications of multiple models, if those models disagree with each other.
Hence, we seek for means to resolve such conflicting interpretations when it comes to optimizing a software system for multiple objectives.

\begin{figure*}[t!]
\centering
\includegraphics[width=\linewidth]{ 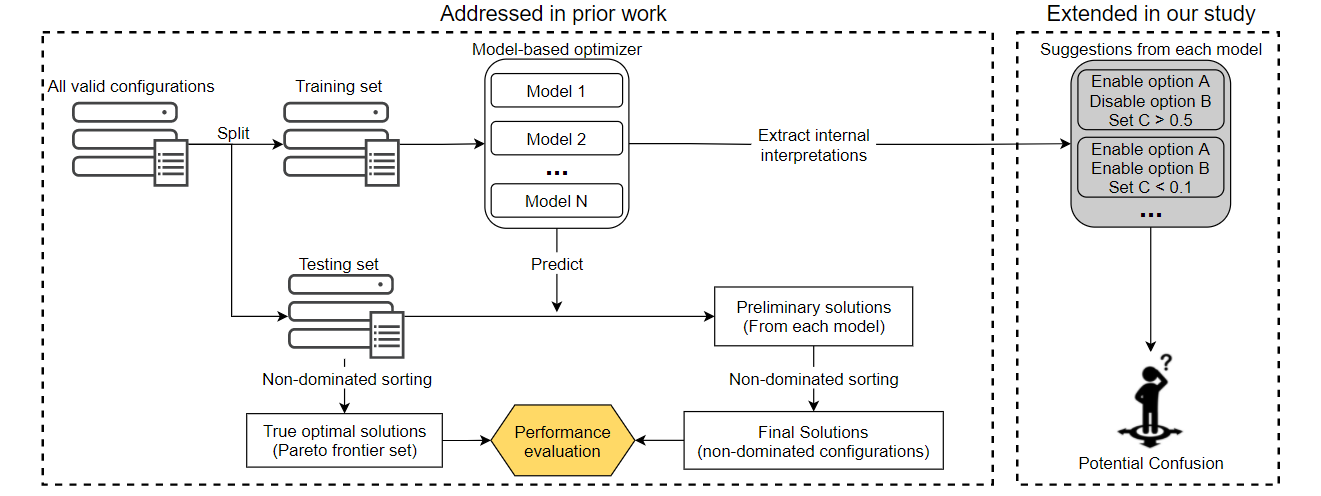}
\caption{An overview of the configuration optimization problem. Prior research focused on the performance of the model, as shown in the bounded region. What has been missed is the evaluation on the inside rationales of the surrogate models.
Prior model-based optimizers only assess the quality of selected solutions (with the gray block being neglected). As shown later in \S\ref{mdp}, models trained on different objective can often disagree on how to optimize a configuration.} 
\label{fig:top}
\end{figure*}

We are concerned with transparency since, as shown in Figure~\ref{fig:top}, ML models that chase different goals might make different recommendations. Therefore, even if each of the machine learners is interpretable, it remains possible that the internals of different machine learners are disagreeing or conflicting with each other: For example, one model might offer an insight that increasing certain option $X_{1}$ can optimize an objective $G_{1}$, while another model believes increasing such option will harm another objective $G_{2}$.  In Table~\ref{tab:example1}, we present some examples of such disagreement observed from dataset explored in this paper.

\subsection{What is a good interpretation?}

Looking through the literature, we can find very little on how to handle model disagreement in multi-objective optimization tasks.
The closest thing we have found to our work is the  Clafer visualizer \cite{antkiewicz2013clafer} environment.
In Clafer, when there is a trade-off among multiple objectives in an optimization problem, the users are asked to make the decision on the trade-off. 

While Clafer improves practitioners' understanding by visualizing the performance of different configuration candidates, such form of interpretations still has its limitations.
Firstly, Clafer's interpretations on how to optimize a configuration task is {\it instance-based}: While candidate configurations with various trade-off among multiple objectives are shown to users, no further insights are generalized by Clafer. In other words, users may be able to obtain a vague picture of what configuration options have positive influence on certain objective (in terms of performance measured), yet those options can usually be conflicting due to competing objectives (see examples in Table~\ref{tab:example1}). In the end, there still lacks of conclusive agreement on which options should be preferred.

Secondly, the problem with human-centered approaches such as the Clafer visualizer is that a repeated empirical result illustrates that humans are very poor oracles for what best improves a project (e.g., in Devanbu's ICSE'16 study \cite{devanbu2016belief} on 500+ developers at Microsoft, even when developers work on the same project, they mostly make conflicting and/or incorrect conclusions about what factors most affect software quality; similar patterns were observe in Shrikanth's ICSE-SEIP'20 study \cite{shrikanth2020assessing}). Hence we seek methods that remove {\em as much as possible} those competing recommendations.

Regarding {\em as much as possible}, sometimes objectives are {\em inherently opposed} in their recommendation direction, due to the semantics of the domain. If this effect has the majority case,  then the method of this paper would be doomed to fail.
That said, the novel result of this paper is that at least for the data sets studied here, the objectives are not {\em inherently apposed}. Since we can generate a disagreement-free model to provide solutions that perform as well or better as those seen produced by other methods, this result should prompt much future work exploring emergent simplicity in multi-objective reasoning in SE.

\begin{table*}[!t]
\centering
\scriptsize
\caption{Configurable software systems used in this paper. The abbreviations of systems are ordered by the total number of valid configurations $|\mathcal{C}|$. The ``B/N'' in the third column indicates the number of binary options and numerical options. The last column represents the number of objectives $O$ in the corresponding project.}
\begin{tabular}{lllllll}
\toprule
Name    & Abbr.        & Domain                   & \#Options (B/N) & $|\mathcal{C}|$   & Performance Measures          & $|{O}|$  \\ 
\midrule
HSQLDB          &SS-A  & SQL database             & 15/0            & 864   & run time, energy, cpu load   & 3\\ 
MariaDB         &SS-B  & SQL database             & 7/3             & 972   & run time, cpu load           & 2\\ 
wc-5d-c5        &SS-C  & streaming process system & 0/5             & 1\,080  & throughput, latency        & 2\\
VP8             &SS-D  & video encoder            & 7/4             & 2\,736  & run time, energy, cpu load & 3\\ 
VP9             &SS-E  & video encoder            & 9/3             & 3\,008  & run time, cpu load         & 2\\ 
rs-6d-c3        &SS-F  & streaming process system & 0/6             & 3\,839  & throughput, latency        & 2\\ 
lrzip           &SS-G  & compression tool         & 9/3             & 5\,184  & run time, cpu load         & 2\\ 
x264            &SS-H  & video encoder            & 17/0            & 4\,608  & run time, cpu load         & 2\\ 
MongoDB         &SS-I  & No-SQL database          & 13/2            & 6\,840  & run time, cpu load         & 2\\ 
LLVM            &SS-J  & compiler                 & 16/0            & 65\,536 & run time, cpu load         & 2\\ 
ExaStencils     &SS-K  & Stencil code generator   & 4/6             & 86\,058 & run time, cpu load         & 2\\ 
\bottomrule
\end{tabular}
\label{tab:CSS}
\end{table*}

\section{Details, Definitions  and Algorithms}\label{Background}

This section refines the motivation of this paper by defining the configuration optimization problem and reviewing prior works associated with the problem. 
Table \ref{tab:CSS} shows the datasets used in our experiments.

\subsection{Configurable Software System (CSS)}

In general, a configurable software system (CSS) contains a set of valid configurations $\mathcal{C}$. Let $c_i \in \mathcal{C}$ represent the $i$th valid configuration and $c_{i,j}$ represent the $j$th option in that configuration. 
In our case studies, the option $c_{i,j}$ is either a numerical parameter or a Boolean value. 
A numerical parameter (i.e., page size) has multiple different numerical values and Boolean values are indicating a certain option as enabled or disabled. 
A configurable software system also contains a set of performance measures $Y$ (i.e., response time, energy consumption, etc.), where $y_{i,k} \in Y$ represents the $k$th performance value of the $i$th configuration in $C$. 
Since we consider configurable systems with more than one performance measure, $k$ is always greater than 1. 
The configuration space $\mathcal{C}$ is referred to as an independent variable while the performance measure space $Y$ is referred to as a dependent variable (i.e., depends on $\mathcal{C}$).
Theoretically, to find an optimal configuration for a software system, we learn the relationship between $\mathcal{C}$ and $Y$ by approximating the function $f: \mathcal{C} \mapsto Y$ that maps configurations onto the performance (objective) space by $f(c_{i,0}, ... ,c_{i,j}) = (y_{i,0}, ..., y_{i,k})$. 
In real-world practice, the evaluations required to approximate the function $f$, also referred to as measurements, are usually expensive and time-consuming. 
Therefore, in this paper, the cost of a configuration optimizer is referred to as the total amount of measurements taken to generate the optimal solutions (may not be actually ``optima'', but one that an optimizer determines to be optimal).

While there always exists an optimal solution in  a single-objective optimization task (where a better solution can be determined by a single criterion), there might not be any optimal solution in a multi-objective optimization task. This might be the situation where no configuration is best in all objectives.
% In a single-objective optimization task (where a better solution can be defined and determined by a single criterion), an optimal solution always exists.
% But often, systems seek to optimize on more than one performance indicator (since different stakeholders may possess distinct expectations and requirements on the software product).
% The problem here is that there might not be any optimal solution in a multi-objective optimization task. 
Therefore, we need to quantify and evaluate the overall quality of a configuration in a different manner, called the {\em domination} relationship. There are two major types of domination relationship: binary domination~\cite{holland1992genetic}, and continuous domination~\cite{laumanns2002combining}. According to the definition of binary domination, a configuration $c_1$ is binary dominant over $c_2$ iff:
\begin{equation}\label{b-dom}
\begin{gathered}
    y_{1,i} \leq y_{2,i} \forall i \in \{ 1, 2, ..., n \}     \text{ and} \\
    y_{1,j} < y_{2,j} \text{ for, at least, one } j \in  \{ 1, 2, ..., n \}     
\end{gathered}
\end{equation}
In contrast, continuous domination defines that a configuration $c_1$ is dominant over $c_2$ iff:
\begin{equation}\label{c-dom}
loss(c_1, c_2) < loss(c_2, c_1)
\end{equation}
where the loss function is defined as:
\begin{equation}\label{loss}
\begin{gathered}
loss(c_1, c_2) = \sum_{j=1}^{n} -e^{(y_{2,j}-y_{1,j}) }\times \frac{1}{n}\\
\text{where } y_{i,j} \text{ is min-max normalized}
\end{gathered}
\end{equation}

Note that in both equations above, the default setting is that the lower performance measure $y$ is preferred.
% In case a certain performance measure needs to be maximized, one can simply reverse the domination relationship returned by the equations above. 
By definitions, a best (optimal) configuration is one that is not dominated by any other configuration, denoted as a non-dominated solution. And the set containing all non-dominated solutions is called Pareto frontier set. In short, the goal in a multi-objective optimization problem is to find as many optimal (non-dominated) solutions as possible while minimizing the evaluation cost (using fewer measurements).

\begin{algorithm}[t!]
\caption{Non-dominated Sorting (NDSorting) \label{algo2}}
% \small
\KwData{$C$ contains configurations await to be sorted;
performance function $f$ maps configurations to the corresponding performance values.
}
\KwResult{A set of non-dominated configurations $C_{nd}$, also referred to as the Pareto frontier.}
\Begin{
    $C_{first} \gets$ $C$\\
    $C_{nd} \gets$ $C$\\
    \While{$C_{first} \neq \emptyset$}{
        $c_1 \gets C_{first}$.pop()\\
        $C_{second} \gets$ $C_{first}$\\
        \While{$C_{second} \neq \emptyset$}{
            $c_2 \gets C_{second}$.pop()\\
            // using either domination as defined in Eq. \ref{b-dom} or Eq. \ref{c-dom} \\
            \If{$f(c_1)$ dominates $f(c_2)$}{
                $C_{nd}$.remove($c_2$)\\
            }
            \If{$f(c_2)$ dominates $f(c_1)$}{
                $C_{nd}$.remove($c_1$)\\
            }
        }
    }
    return $C_{nd}$
}
\end{algorithm}

To compute the non-dominated solution set, the
{\bf non-dominated sorting} process is required, which has a runtime complexity of $O(MN^3)$ where $M$ is the number of objectives and $N$ is the population size. It is also noteworthy that NSGA-II~\cite{deb2002fast} proposed a fast and elitism sorting approach that reduced this complexity to $O(MN^2)$. As described in Algorithm~\ref{algo2}, this paper follow the same sorting process used in NSGA-II.
% It is apparent that the execution time will increase dramatically as the configuration space scales up. 
This is important since our approach provides a faster workaround that only uses non-dominating sorting during model training and avoids the use of it when executing the trained optimizer.

\subsection{Sequential Model-Based Optimization (SMBO)}\label{smo}

Previous work has commented on the cost of performing configuration exploration. Given all the possible configurations, it can be prohibitively expensive and time consuming to run them all.

\begin{algorithm}[!t]
\caption{SMBO (e.g., FLASH)\label{algo1}}
% \small
\KwData{$C$ contains all candidate samples;
% $C\_train$ keep track of all configurations evaluated so far;
performance function $f$ maps configurations to the corresponding performance values;
initialized surrogate models $M$ contain multiple models, one per objective; $budget$ denotes the stopping criterion; $S$ denotes the non-dominated solutions found so far.
}
\KwResult{Optimized models $M$, non-dominated configurations $C_{nd}$}
\Begin{
$C_{train}$ $\gets$ Random$(C)$ // initialize training samples\\
$C$.remove($C_{train}$) \\
$M \gets$ FitModel($C_{train}$,$f$)\\ 
$S \gets \emptyset$ \\
\While{$budget \geq 0$ }{

    $C\_new \gets$ SelectConfig($M, f, C$)\\
    $C$.remove($C\_new$)\\
    $C\_train$.add($C\_new$)\\
    $M$.train($C_{train}, f$)\\
    $C_{nd} \gets$ NDSorting($C_{train}, f$) // as described in Algorithm~\ref{algo2}\\
    \eIf{$C_{nd}$.isUpdated}{
   $continue$\\
   }
   {
   $budget \gets budget-1$ 
  }
%   $i \gets i+1$
}
return $M$, $C_{nd}$
}
\end{algorithm}

In the AI literature, one method to explore
a large and complex space without excessive
sample is sequential model-based optimization (SMBO).
SMBO is an efficient tactic to find extremes of a performance (objective) function that is expensive to evaluate (in terms of measurement cost and time). Also referred to as Bayesian optimization in literature, SMBO can better incorporate prior knowledge in the form of already measured solutions (in our case, configurations) as compared to traditional optimization algorithms \cite{brochu2010tutorial, snoek12gp}. By sequentially updating and learning from the prior knowledge, SMBO can make estimations on the rest of unmeasured solutions so that it can locate the most ``interesting'' (estimated to have better performances) space for further sampling.

As shown in Algorithm \ref{algo2} and Algorithm \ref{algo1}, the concept of SMBO is simply and straightforward: {\em Given the current knowledge learned about the problem space, where should the procedure explore next?}
The advantages of SMBO over other traditional MOEA approaches (e.g., NSGA-II \cite{deb2002fast}, SPEA2 \cite{zitzler2001spea2}, MOEA/D \cite{zhang2007moea}) are:
\begin{itemize}
    \item SMBO  explores the unknown part of configuration space sequentially based on knowledge already gained from the optimization so far. This results in much fewer evaluations required to achieve the termination criteria (e.g., only 70 samples needed to explore a space of nearly 80,000 configurations, while traditional genetic algorithms require much more evaluations\footnote{Holland's advice~\cite{holland1992genetic} for genetic algorithms (such as  NSGA-II and MOEA/D) is that 100 individuals need to be evolved over 100 generations; i.e., $10^4$ evaluations in all. }).
    \item SMBO contains a set of surrogate models, on which the optimization is performed. Each model is fitted for a unique objective. After the termination, the surrogate models can provide human-comprehensible insights on how to achieve better performance for different objectives.
\end{itemize}
It is undeniable that traditional multi-objective optimizers still have their values, especially when the valid search space is vast and the evaluation of solutions is inexpensive. Unfortunately, the problems explored in this paper do not fall into this category. Configurable software systems can often contain constraints among configuration options, which can reduces the valid configuration space vastly~\cite{kaltenecker2020interplay}. For example, the system SS-H in Table \ref{tab:CSS} has 17 binary options yet with only 4\,608 valid configurations in total. That is, the ratio of valid solutions in the whole search space is $4608/(2^{17}) = 3.5\%$. Under such circumstance, it is believed that guidelines should be adopted to improve cost efficiency of sampling~\cite{sarkar2015cost}. Therefore, we believe that SMBO is a more suitable approach for our problem case.

\subsection{Finding Interpretable SMBO (with FLASH)}
Many approaches have been proposed using the SMBO framework.  In terms of transparency, Nair et al.'s FLASH system~\cite{nair2018finding}, is somewhat unique
in that if offers a succinct summary of
the learned model~\cite{9463120}.  As we shall see,
this directly addresses
one problem (model transparency) but
introduces another (model disagreement).

First proposed by Nair et al., FLASH can achieve on-par performance while overcoming the shortcomings of prior SMBO methods: Nair et al. reported that FLASH takes $10^{2}$ evaluations-- which is much less  than the $10^{4}$ evaluations required by other optimizers. Such improvement is notable since it makes FLASH more scalable to large-scale systems with a vast search space. For example, the data used in this paper required 6 calendar months to collect (running on a multi-core CPU farm). While that data
is necessary to certify a new algorithm (like VEER), once that algorithms is fielded, it needs to respect the practical difficulties associated with data collection.

While Gaussian process models (GPM) are often used~\cite{golovin2017google} in SMBO,
Nair et al found that, for optimizing configurable software, GPM scales very poorly to larger dimensional data~\cite{nair2018finding}. 
Nair et al. found that a faster, and the more scalable, system can be implemented using regression trees. In FLASH, each objective is modeled as a separate Classification and Regression Tree (CART) model.
They found that even if regression trees  are somewhat incorrect about their predictions,
those approximate predictions can still be used to rank different candidate configurations~\cite{nair2017using}.
FLASH is implemented following the general SMBO framework as described in Algorithm \ref{algo1}, where the surrogate models $M$ are CART learners.

\begin{figure}[t!]
\centering
\includegraphics[width=.75\textwidth]{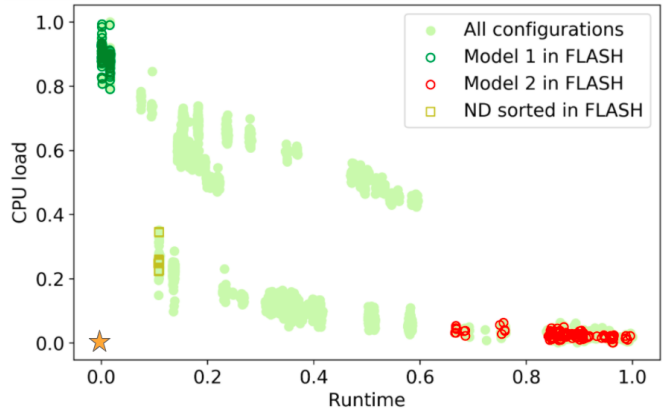}
\caption{Visualization of the model disagreement problem in FLASH~\cite{nair2018finding}. The internal rationales of the surrogate models are as presented in Table. \ref{tab:example1}.  ``ND" means non-dominated. The x and y axis represent the two performance objectives (min-max scaled) in the dataset SS-E in Table \ref{tab:CSS}. The star at the bottom-left corner indicates the ideal optimum. 
% It is notable that the final solutions after non-dominated sorting are distant from solutions selected by both single-objective surrogate models.  
} 
\label{fig:mdp}
\end{figure}

There are several reasons we chose FLASH:
\begin{itemize}
    \item Nair et al. showed that
    FLASH can handle models orders of magnitude 
    faster than a prior state-of-the-art methods based on Gaussian process models~\cite{zuluaga2016varepsilon}. 
    \item FLASH makes its conclusions after very few samples to the domain.
    \item  Due to the small size of the sample space,
    then the decisions used by FLASH generate very small models (one per objective). Hence, FLASH can produce the human-readable models needed
    for the AI transparency issues discussed
    in \S\ref{rw}
\end{itemize}

\subsection{Model Disagreement Problem and FLASH}\label{mdp}

For our purposes,
FLASH is both a success and a failure.
Firstly, it fixed the scalability issues of GPM.
At the same time, it turns out that model disagreement is rampant in
the models generated by FLASH
(or example,
see all the examples of disagree in   Table~\ref{tab:example1} were generated  by FLASH).

To understand the root cause of such model disagreement, we attempted to visualize the inside rationales of the two surrogate models. As shown in Figure \ref{fig:mdp}, the ``optimal'' solutions selected by the two models are rather different, which is totally reasonable and expected, given that they are optimizing for different objectives. However, it is noteworthy that the final ``optimal'' solutions yielded by FLASH (which is selected by the non-dominated sorting procedure) share no similarity with either of the two solution sets. That is to say, the interpretations generated by the two models alone are not ``final'', and if users merely rely on such interpretations to locate optimal configurations, they are more likely to obtain sub-optimal solutions that are distant to the ones yielded by FLASH. Since the non-dominated sorting procedure is a non-parametric process from which we cannot extract interpretations, we need an additional model that can mimic the performance of the sorting procedure meanwhile allowing us to obtain comprehensible insights.

\section{VEER:  Disagreement-free Multi-objective Optimizer}\label{veer}
As a response to our insights in \S\ref{mdp}, we design VEER based on the following design choices:
\begin{itemize}
    % \item To ensure our interpretations are interaction-aware, we use CART as our surrogate model because the internals of CART can be presented as decision rules, which are highly readable and understandable to users.
    % \item To ensure our interpretations are generalizable, we choose to generate model-wise, rather than instance-wise, interpretations. 
    \item To ensure our model provides final interpretations, we design a method to reduce the multi-dimension objective space into a single-dimensional space. This will enable us to provide interpretations that take into account the overall performance across multiple objectives. 
    \item To ensure our interpretation is confusion-free, we use one single-output model as the new surrogate model. This way, we avoid the dilemma that the same candidate solution (configuration) gets ranked differently by different learners (or different outputs from one multi-output model).
    \item To conduct a fair comparison with FLASH, and to obtain rule-based interpretations, we choose to use CART as the new surrogate model to optimize on the synthetic single-dimension space. In future deployment, VEER is applicable to any interpretable models such as Linear Regression and Naive Bayes.
    % \item Apart from training multiple learners on different objectives separately, VEER will further synthesize a hyper-space using the heuristic named ZIGZAG to "compress" multiple objectives into single-dimension space. 
    % % As a result, VEER will return one single-output learner as the final model that serves to generate the configuration solution set. 
    % \item In order to offer comprehensible interpretations about feature interactions, CART\cite{breiman1984classification} is selected as the surrogate model to be trained on the synthetic hyper-space.  
\end{itemize}

\begin{figure*}[b!]
\centering
\includegraphics[width=.99\linewidth]{ 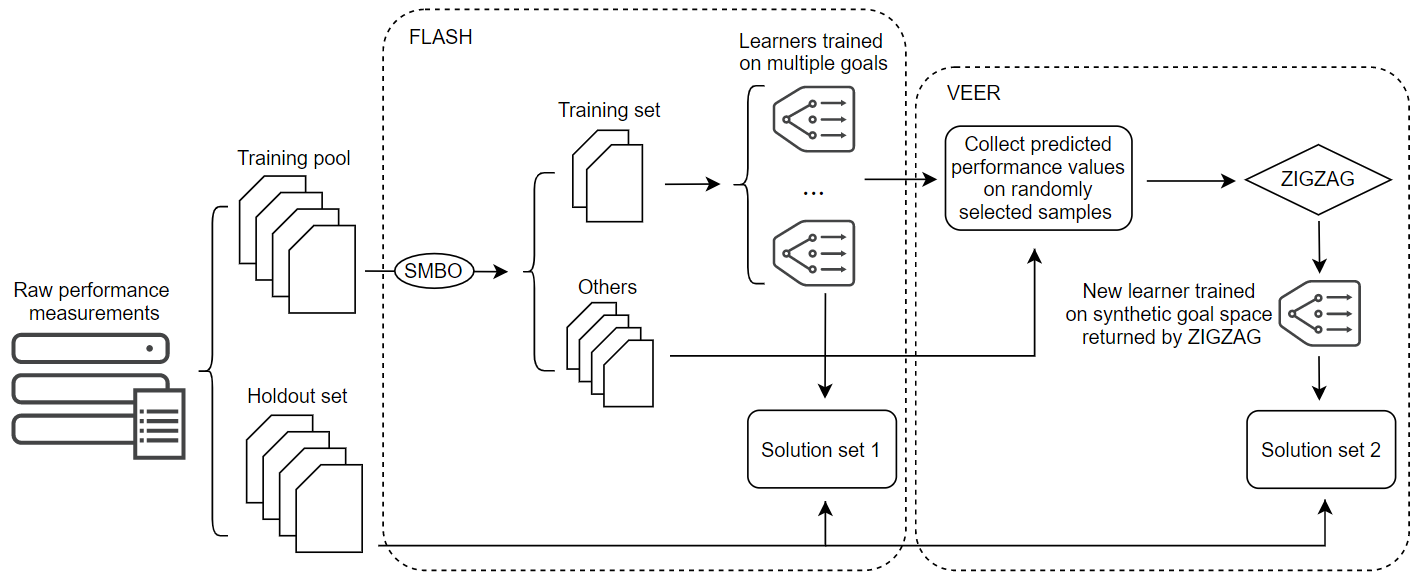}
\caption{An overview of the framework using VEER, and the evaluation setup used in our experimentation. Note that the first block is described in Algorithm~\ref{algo2} and Algorithm~\ref{algo1}, whereas the second block, VEER, is described in Algorithm~\ref{algo3}.} 
\label{fig:framework}
\end{figure*}

\begin{figure}[t!]
\centering
\includegraphics[width=.85\textwidth]{ 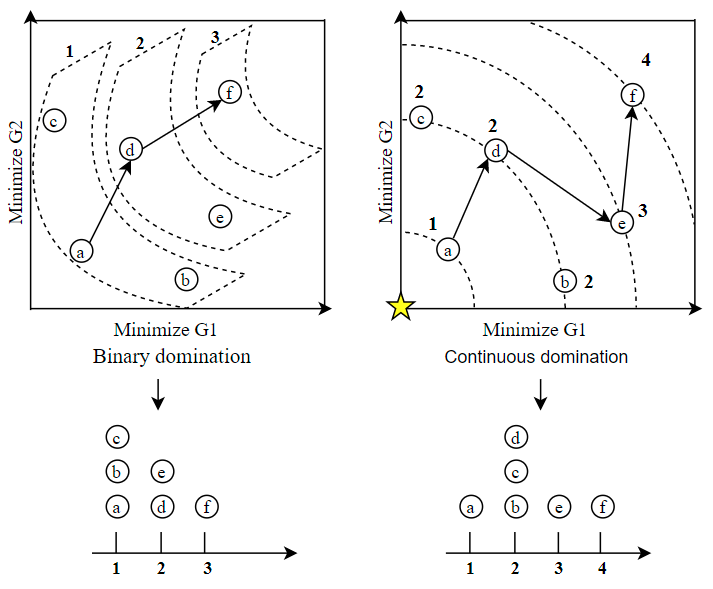}
\caption{ZIGZAG: candidate configurations are ranked according to their ability of dominating other configurations across the configuration space. 
Starting at the best objective (bottom left, which is ranked \#0), VEER  zigzags around objective space looking for the next best unvisited objective. Configurations that cannot dominate or be dominated by each other are assigned the same rank.} 
\label{fig:zigzag}
\end{figure}

An overview of the framework using VEER is shown in Figure \ref{fig:framework}
(which also includes the experimental rig used in this paper). The component ZIGZAG, as illustrated in Figure \ref{fig:zigzag}, is the core heuristic used in VEER to generate the single-dimension objective space, and its implementation varies for different definitions of domination. 
In this paper, we choose to use the heuristic implemented with continuous domination because we believe it can better reflect and preserve the domination relationship among solutions in our setting: As one can observe from the two examples in Figure \ref{fig:zigzag}, when using binary domination, point $c$ is assigned a higher rank than point $e$. However, if we only look at these two solutions, neither of them can dominate each other, thus, assigning them different ranks seems less reasonable. Such tricky situations can be avoided in the continuous domination scenario because points are ranked precisely according to their distance toward the "heaven" point (where both objectives are optimized). 

\begin{algorithm}[t!]
\caption{VEER \label{algo3}}
% \small
\KwData{$M$ is a set of SMBO models already trained. $M_i \in M$ corresponds to the models trained for the $i$th
 objective. $C$ contains all valid configurations.}
\KwResult{$M_{VEER}$ is the resulting model that will be utilized to replace the original $M$ during deployment. }
\Begin{
    $C_{sample} \gets$ RandomSample($C$, ration)\\
    $Y \gets \{\}$ \\
    $PR \gets \{\}$ \\
    \For{$M_i \in M$}{
        $y_i \gets M_{i}$.predict($C_{sample}$)\\
        $Y$.append($y_i$) // record the predicted performance values for the $i$th objectives\\
    }
    $i \gets 1$\\
    \While{$C_{sample} \neq \emptyset$}{
        $C_{nd} \gets$ NDSorting($C_{sample}, Y$)\\
        $C_{sample}$.remove($C_{nd}$)\\
        $PR$.update($C_{sample}, i$) // record the current Pareto rank as depicted in Figure~\ref{fig:zigzag}\\ 
        $i \gets i+1$\\
    }
    $M_{VEER} \gets$ FitModel($C_{sample}, PR$)\\ // Fit a new model using PR as the dependent feature.\\
    $C_final \gets argmin(M_{VEER}$.predict($C$)$)$\\
    //Those predicted to have lowest PR will be returned as final choice of configurations.\\
    return $M_{VEER}, C_{final}$
}
\end{algorithm}

VEER uses CART as the final surrogate model motivated by effectiveness and interpretability considerations. As previously shown by FLASH, an optimizer model built with CART can achieve comparable and sometimes superior performance in multi-objective optimization as compared to GPM. One reason that CART scales better than GPM in data of larger dimensionality is that CART does not presume the "smoothness" of the data space: models like GPM make an assumption to the data space that configurations closer to each other have similar performance. Such assumption can usually be invalid because a seemingly small change in configuration options might actually represent a crucial shift in configuration strategy (i.e., the choice of data structure often has a substantial impact on the performance of a storage-oriented system). Our second consideration is interpretability. Gigerenzer pointed out that tree-structured models express great rationality and interpretability, making it easier for users to obtain actionable insights about the data \cite{gigerenzer2008heuristics}. A recent survey about practitioners' beliefs about visual explanations of defect prediction models also shows that CART is favoured by practitioners for its interpretability \cite{jiarpakdee2021practitioners}. The result of the survey reported that among 6 methods of offering visual explanations, CART is ranked as the 1st tier in terms of insightfulness and quality of the generated visual explanations. 

Algorithm~\ref{algo3} offers details on {\IT}'s internal workings.

\section{Research Questions}\label{rqs}
This section illustrates research questions and our strategy to address these questions.

To systematically evaluate its merits, we compared VEER with one of the most recent state-of-the-art configuration optimizer, FLASH. Moreover, to verify whether the model disagreement problem can be tamed via a simple and naive approach, we also implemented two variants of FLASH to serve as the baseline. 

Our research questions are geared towards assessing the performance of VEER regarding 3 aspects: (a) \textbf{effectiveness} of configuration solutions generated from the model using VEER heuristic, (b) \textbf{interpretability} of VEER, and the (c) \textbf{execution time} of the model using VEER heuristic.
More specifically, we ask the following research questions:

\textbf{RQ1:} \textit{Is there a model disagreement problem for multi-objective optimization of CSS?}

\noindent
Before doing anything else, we need to first motivate the investigation of this paper. In this research question, we ask whether the standard model-based method (FLASH) used in the multi-objective configuration problem results in conflicting suggestions on optimizing different objectives. To measure the level of such conflicts, we used the rank correlation measurement called Kendall's $\tau$ test to assess whether learners trained on different objectives rank the candidate configurations in different orders.
% We will find that confusing and contradictory interpretations
% arise when configurations believed to have good performance in one objective are often found to have poor performance in other objective(s).  

\textbf{RQ2:} \textit{Is the disagreement problem ``linearly solvable''? }

\noindent 
All the technology proposed in this paper is superfluous if an existing alternate method can handle the problem of interpreting multi-objective optimization results. 
% For example, a {\em multi-objective regression tree learner} builds one tree that optimizes for minimizing the linear sum of the  normalized
% error of all the objectives. Such learners are readily available (see class MultiOutputRegressor in scikit-learn~\cite{scikit-learn}). Another approach is to replace the multiple objectives with their weighted sums. There are various weighted sum equations, and one of the major issues of this approach is that it assumes the convexity of the optimization problem: also shown later, in cases where there might be a non-convex optimal solution set, the performance could be largely compromised.
Therefore, we attempt to edit the default FLASH into two variants, using two different heuristic function to reduce the objective space: (a) the weighted sum equation and one single-output regression learner, or (b) a multi-output regression learner. The two benchmark methods are referred to as {\bf SingleWeight} and {\bf MultiOut} respectively.
Our experiments will show that both methods have significant shortcomings.

\textbf{RQ3:} \textit{Can VEER resolve the model disagreement problem while maintaining on-par performance with benchmark methods?}

\noindent 
% Given that we have shown that the feature interaction problem exists (in {\bf RQ1}) and that it cannot be readily solved
% by widely available  methods (in {\bf RQ2}), we now need to check if the proposed technology of this paper is effective.
Here, we test  whether introduction of the VEER heuristic will compromise performance of the state-of-the-art optimizer . To measure that, we evaluate not only the quality of the returned solutions, but also the robustness of the model. Measurements used to evaluate the merits of VEER and other benchmark methods are elaborated in \S\ref{criteria}

\textbf{RQ4:} \textit{Can VEER reduce the execution time?}

\noindent We explore other positive side effects brought by VEER. One of the most apparent improvements of VEER is that since multiple learners within the optimizer are replaced by a single learner, the model can now skip the step of computing domination sort across the whole holdout set, which is computationally very intense. To assess the extent of the improvement, we will record the execution time of VEER when applying it to the holdout set, and compare it with that of other benchmark methods. 

\section{Experimental Setup}\label{experiment}
To answer our research questions, our experiment compares the performance of configuration optimizers using VEER heuristic against those using alternative heuristics. 

As depicted in Figure \ref{fig:framework}, all the configuration optimizers in our experiment will divide the configuration space into 2 sets: the training pool, and the holdout set (we split them 50\% to 50\%). Each optimizer sequentially samples along the training pool to add the selected next most informative configuration item into the training set (and this set is a subset of the training pool). The selected configurations and the corresponding performance measures are then used to train the machine learners within the optimizer. 
Then, the performance of each optimizer is evaluated using the holdout set. Apart from that, VEER has one additional step, which is to use the ZIGZAG process as illustrated in Figure \ref{fig:zigzag} to train a hyper-space model using random samples chosen from the rest of the training pool. Note that, since this step does not require to access the actual performance measures $Y$ of the additionally chosen samples $C$, it will not increase the measurement cost in the real-world application.
Finally, to assess the stability and reliability of our approach in a statistical manner, our experimentation will randomly choose 50\% of the configuration space as the holdout set. To reduce the effect of the random seeds, the whole process is repeated for 100 times. 
For replication purposes, our code and datasets are on-line: \url{https://github.com/anonymous12138/multiobj}.

\subsection{Data}
To empirically evaluate the effectiveness of our approach, we use datasets collected from different configurable software systems. 
Each dataset contains the whole population of all valid configurations of that system and the performance measures of each configuration (by ``all'', we mean all combinations given the selected configuration options). 
Table \ref{tab:CSS} describes the nature of each dataset. 
We selected the datasets based on the following criteria: (1) different sizes to examine the scalability and robustness, (2) different domains to improve external validity, and (3) different application domains (client-server and desktop) to cover different performance aspects (i.e., run time of compressing a video vs. run time to perform a set of actions on a database).
Among all the datasets, SS-C and SS-F are datasets used in FLASH. 
Others (SS-A, SS-B, SS-D, SS-E, and SS-G to SS-K) are datasets recently collected by us.
For datasets that are collected by us, we applied the following process:
To reduce measurement noise, we executed all measurements in isolation (i.e., no other tasks are performed) on machines with minimal Debian 9 or Debian 10 installations and repeated each measurement 3--5 times. 
While the standard deviation of a performance measure exceeded $10\%$, we repeated the measurement of the configuration. 

\subsection{Baselines}
To assess our approach comprehensively, we included several SMBO methods as benchmark methods in out experiment. At first, we collected some existing open-source SMBO methods, such as FLASH \cite{nair2018finding}, HyperOpt \cite{bergstra2013hyperopt} (using TPE \cite{bergstra2011algorithms} as surrogate models) and SMAC \cite{hutter11smac}. Unfortunately, we found that both HyperOpt and SMAC do not support customized constraints on the search space. As illustrated in \S\ref{smo}, such constraints are rather crucial as they filter out over 99\% of invalid configurations~\cite{kaltenecker2020interplay}. Hence,
we had to  implement our own versions of the
  {\bf SingleWeight} and {\bf MultiOut} 
alogorithms  described in \S\ref{rqs}. The implementation follows the SMBO framework as described in prior works \cite{bergstra2011algorithms, nair2018finding, brochu2010tutorial}, denoted as {\bf SingleWeight} and {\bf MultiOut}\footnote{And the source code for that implementation can be found
in the reproduction package mentioned in our abstract}. The major difference among these benchmark methods is the output format of surrogate models (either {\em single-output} or {\em multi-output}) and option of using either  {\em weighted sum} or {\em non-dominated sorting} to select the final solutions. The implementation of all benchmark methods is available in our online repository.

% As mentioned above, FLASH uses standard regression trees (the CART algorithm~\cite{breiman1984classification}) while FLASH\_naive2 uses MultiOutputRegression tree. In our experiments, we use CART with default parameters.
% One reason to use regression trees, as opposed to (say) Deep Learning or instance-based nearest neighbor algorithms
% is that the regression tree generates a high-level human interpretable  model~\cite{chen2018applications}. As shown later, such highly interpretable models are also helpful when we attempt to diagnose the internals of a model.

% \begin{figure}[t!]
% \centering
% \includegraphics[width=.35\textwidth]{ gd.png}
% \caption{An example illustrating how generational distance (GD) is calculated. The GD for a single predicted non-dominated solution is the distance between it and the nearest actual non-dominated solution (denoted as solid lines in the figure). The aggregation of all the actual non-dominated solutions is known as the Pareto Frontier solution set. The GD of a solution set is the mean value of GDs of all solutions within the set.} 
% \label{fig:gd}
% \end{figure}

% \begin{figure}[t!]
% \centering
% \includegraphics[width=.35\textwidth]{ kendall.png}
% \caption{A visualized example illustrating how Kendall's $\tau$ coefficient in Eq. \ref{tau} is calculated. As indicated by the coefficient, the 2 lists of ranks assigned to the same set of variables are positively correlated. } 
% \label{fig:kendall}
% \end{figure}

\subsection{Performance Criteria} \label{criteria}
% In this section we will introduce the criteria used to address research questions asked in \S\ref{evaluation}.
First of all, to assess \textbf{effectiveness} of our approach against benchmark models, we choose to measure the quality of the solutions set returned by each model. A solution set contains configurations that a model believes to have optimal performance among all configurations. Given the definitions of binary and continuous domination, a solution set can contain more than one configuration. 
To measure the quality of the solution set, this paper uses generational distance \cite{van1999multiobjective} (GD) as the indicator. GD computes the average distance between the solution set returned by a model and the actual optimal solution set. There are other indicator such as inverted generational distance \cite{coello2004study} (IGD)  and hypervolume \cite{huband2003evolution} (HV). However, prior research suggests GD as a more suitable metric to uniformly reflect the overall quality of the solution set~\cite{agrawal2020better}.
For example, if one intentionally adds very poor solutions into the solution set, IGD and HV cannot reflect the change in the overall quality of the new solution set.

Secondly, to assess the \textbf{interpretability} of a model, in this paper we are specifically assessing the level of \textbf{disagreement} among multiple learners within a multi-objective model.
We argue that if interpretations extracted from different learners are conflicting or disagreeing with each other, the model will fail to provide stakeholders with unequivocal insights that are truly informatively or actionable. 
% That is, when looking at configurations return by one model trained on an objective $G_A$, one should expect these configurations to have a similarly good performance on another objective $G_B$. Otherwise, an opposite scenario would indicate that feature interactions used to locate optimal configurations on $G_A$ are different from interactions used to locate optimal configurations on $G_B$.  
In that spirit, we use a rank correlation test, Kendall's  $\tau$ test \cite{kendall1948rank}, to measure the extent of disagreements among learners built on different objectives.  Kendall's  $\tau$ test is a non-parametric statistical test which can be used to measure the ordinal association between 2 lists of measured variables (in this paper, 2 objective values on the same configurations). According to the definition from Kendall correlation, we first categorize any pair of configurations by their performance measure into 2 kinds: discordant pairs and concordant pairs. Let $(A_i,B_i)$ and $(A_j, B_j)$ denote a pair of configurations, represented by the performance measures in objective $A$ and objective $B$. This pair is concordant if the sorted order of  $(A_{i},A_{j})$ and $(B_{i},B_{j})$ agrees: one configuration has better performance than the other in both objectives\footnote{We define a concordant pair in tasks of more than 2 objectives in a similar manner: one configuration has better performance than the other in all objectives. This is not originally defined by Kendall, but we believe it is a proper extension.}. Otherwise, the pair is discordant. After that we compute the Kendall coefficient  $\tau$ using the following equation:
\begin{equation} \label{tau}
   \tau =  \frac{(P - Q)}{(P + Q)} 
\end{equation}
where P is the number of concordant pairs, Q the number of discordant pairs.
In general, the Kendall correlation is high when the 2 variables are ranked similarly, and the correlation is low when the 2 variables are ranked differently. More specifically in our case study, a positive  $\tau$ coefficient means a relatively similar ordering among different objectives, which indicates less disagreement among different learners in a multi-objective model; A  $\tau$ coefficient near 0 means the 2 lists of ranks assigned by different learners have no correlation at all; A negative coefficient means the learners rank configurations in somehow opposite order.
% Fig. \ref{fig:kendall} shows a visualized example of how to assess the rank correlation using Eq. \ref{tau}.

Finally, to assess \textbf{computational complexity} of our approach, we measure the execution time of applying each model on the holdout data to generate a solution set. All the above analyses and measurements were executed on a 64-bit Windows 10 machine with a 2.2 GHz 4-core Intel Core i5 processor and 8 GB of RAM.

% \begin{table}[t!]
% \centering
% \begin{tabular}{lcc}
% \toprule
%  &
%   \begin{tabular}[c]{@{}l@{}}Performance \\  on 1st objective\end{tabular} &
%   \begin{tabular}[c]{@{}l@{}}Performance \\ on 2nd objective\end{tabular} \\ 
%   \midrule
% \begin{tabular}[c]{@{}l@{}}Configurations selected  \\for the 1st objective\end{tabular} &
%   $C_{1}G_{1}$ &
%   $C_{1}G_{2}$ \\ \hline
% \begin{tabular}[c]{@{}l@{}}Configurations selected  \\for the 2nd objective\end{tabular} &
%   $C_{2}G_{1}$ &
%   $C_{2}G_{2}$ \\ 
%   \bottomrule
% \end{tabular}
% \caption{ Confusion matrix used to indicate the extent of conflicting feature interactions in 2-objective tasks. $C_{i}G_{j}$ in each cell means the $j$th objective values of configurations $C_i$ selected by the model trained on $i$th objective.}
% \label{tab:confusion-matrix}
% \end{table}

\subsection{Statistical Analysis}\label{stats}

To make comparisons among all algorithms on a single project, we use
a non-parametric significance
test {\em and} a non-parametric effect size. 
% \respto{2a1} \BLACK
Specifically, we use the Scott-Knott test~\cite{mittas2012ranking} that sorts the list of treatments (in this paper, VEER and baselines) by their median scores. After the sorting, it then splits the list into two sub-lists. The objective for such a split is to maximize the expected value of differences $E(\Delta)$ in the observed performances before and after division~\cite{xia2018hyperparameter}:
\begin{equation}
    E(\Delta) = \frac{|l_1|}{|l|}abs(E({l_1}) - E({l}))^2 + \frac{|l_2|}{|l|}abs(E({l_2}) - E({l}))^2
\end{equation}
where $|l_1|$ means the size of list $l_1$.
% \bi
%     \item
%     $|l|$, $|l_1|$, and $|l_2|$: Size of list $l$, $l_1$, and $l_2$.
%     \item
%     $\overline{l}$, $\overline{l_1}$, and $\overline{l_2}$: Mean value of list $l$, $l_1$, and $l_2$.
%     \item
%     $abs$: Calculates the absolute value.
% \ei
% After the best split is declared by the formula above, Scott-Knott then implements some statistical hypothesis tests to check whether the division is useful or not. Here ``useful'' means $l_1$ and $l_2$ differ significantly by applying hypothesis test $H$. If the division is checked as a useful split, the Scott-Knott analysis will then run recursively on each half of the best split until no division can be made. In our study, 
% the hypothesis test $H$ is the cliff's delta non-parametric effect size measure~\cite{macbeth2011cliff}.  The Cliff's Delta non-parametric effect size test explores two lists $A$ and $B$ with size $|A|$ and $|B|$:
% \begin{equation}
%     \mathit{Delta} = \frac{\sum\limits_{x \in A} \sum\limits_{y \in B} \left\{ \begin{array}{l}
%                     +1, \mbox{   if $x > y$}\\
%                     -1, \mbox{   if $x < y$}\\
%                     0,  \mbox{   if $x = y$}
%                 \end{array} \right.}{|A||B|}
% \end{equation}
The Scott-Knott test assigns ranks to each result set; the higher the rank, the better the result. Two results will be ranked the same if the difference between the distributions is not significant.
In this expression, Cliff's Delta estimates the probability that a value in list $A$ is greater than a value in list $B$, minus the reverse probability~\cite{macbeth2011cliff}.   A division passes this hypothesis test if it is not a ``small'' effect ($Delta \geq 0.147$). 
This hypothesis test and its effect size are supported by Hess and Kromery~\cite{hess2004robust}.

\section{Results}\label{result}
This section provides experiment results that answer the research questions (RQs) previously discussed.

\subsection{RQ1}
\textbf{Is there a model disagreement problem for multi-objective optimization of CSS?}

Table~\ref{tab:rc} checks for the existence of this problem by showing the \textit{rank correlation} of FLASH, which is measured by Kendall's $\tau$ coefficient.
% Ideally, a multi-objective optimizer without any disagreement problem is ought to have a high \textit{rank correlation} among all the objectives. The intuition behind is that: if the feature interactions within different learners are coherent to each other, the same configuration should be ranked in a very similar order by all the learners, which result in a high Kendall's $\tau$ coefficient. 
As shown in Table \ref{tab:rc}, there are a few systems where the \textit{rank correlation} is relatively high (SS-I and SS-K). This could be because the objectives in those systems are not so conflicting. In such cases, feature interactions learned from different learners are likely to be homogeneous since essentially these learners can substitute each other without harming the model performance. 
On the other hand, for most of the case studies in this paper, we do observe that configurations are ranked in a rather opposite way by learners trained on different objectives.   

In summary, we answer \textbf{RQ1} as follows:
\begin{blockquote}
% [width=\linewidth,colbacktitle=gray,title={Result 1}]  
\textbf{Answer 1}: 
     In our case studies, we can assert that multi-objective configurable software systems often have extensive
      interpretability problems, in terms of model disagreement among multiple learners. 
\end{blockquote}

\begin{table}[t!]
\centering
\small
\caption{RQ1 and RQ2 result: Median values of the Kendall's $\tau$ coefficient. Higher coefficients are better, and the lowest score(s) are highlighted in each dataset. Please note that the default FLASH (with no additional heuristics or linear variants) will simply be referred to as FLASH.}
\begin{tabular}{lcccc}
\toprule
\small
 & FLASH (Default) & SingleWeight & MultiOut & VEER \\ \hline
SS-A  & \cellcolor[HTML]{DADADA}{0.47} &{1.00}  &{0.82}       & 1.00    \\
SS-B  & {0.16} &{1.00}  &\cellcolor[HTML]{DADADA}{0.13} & 1.00  \\
SS-C  & \cellcolor[HTML]{DADADA}{0.60} &{1.00} &{0.65}      & 1.00    \\
SS-D  & \cellcolor[HTML]{DADADA}{-0.17} &{1.00} & {-0.07}    & 1.00    \\
SS-E  & \cellcolor[HTML]{DADADA}{-0.56} &{1.00} & {-0.54}     & 1.00    \\
SS-F  & {-0.30} & {1.00} & {-0.30} & 1.00  \\
SS-G  & {-0.47}  &{1.00} & \cellcolor[HTML]{DADADA}{-0.51}  & 1.00 \\
SS-H  &\cellcolor[HTML]{DADADA}{-0.69} &{1.00} & \cellcolor[HTML]{DADADA}{-0.69}  & 1.00  \\
SS-I  & \cellcolor[HTML]{DADADA}{0.73} &{1.00} & \cellcolor[HTML]{DADADA}{0.73}  & 1.00  \\
SS-J  &{-0.08} &{1.00} & \cellcolor[HTML]{DADADA}{-0.17}     & 1.00    \\
SS-K  & \cellcolor[HTML]{DADADA}{0.87} & {1.00}& \cellcolor[HTML]{DADADA}{0.88}  & 1.00  \\
\bottomrule
\end{tabular}

\label{tab:rc}
\end{table}

\subsection{RQ2}
\textbf{Is the disagreement problem “linearly solvable”?}

This section compares the results of FLASH (which works on each objective separately)
to {\bf MultiOut} and {\bf SingleWeight} (which work on some linear combinations of the objectives).

For {\bf MultiOut}, we show in Table~\ref{tab:rc} that we can increase the rank correlation by replacing the multiple single-output CART learners with a single multi-output CART learner. This could be a good sign implying that in some cases FLASH can be easily improved in the model disagreement problem via making a small mutation on its original implementation. However, such improvement does have its upper bound, given the disagreement still exists pervasively among all datasets.

As for {\bf SingleWeight}, because we beforehand transformed multi-objective space into a single objective via a weighted sum function, the optimizer now requires one surrogate model. This totally resolved the disagreement problem by reducing the arity of output.
However, as later reported in Table \ref{tab:rq3}, this approach can sometimes compromise the performance significantly. We conjecture that it could be because the optimal solutions are not uniformly distributed or the optimal solutions reside in a non-convex region which leads the weighted sum approach to fail.  

In summary, we answer \textbf{RQ2} as follows:
\begin{blockquote}
% [width=\linewidth,colbacktitle=gray,title={Result 1}]  
\textbf{Answer 2}: 
     Neither {\bf SingleWeight} nor {\bf MultiOut} can fix the disagreement issue while not risking to comprise the performance. That is to say, this problem is not "linearly solvable". 
\end{blockquote}

\begin{table}[t]
\centering
\small
\caption{RQ3 result: Median values of generational distance (GD) for all 4 methods. Each row
highlights the  GD value(s) that are statistically significantly {\bf\underline{worst}} by more than a small effect size  (as determined by   statistical tests of \S\ref{stats}).}
\begin{tabular}{lcccc}
\toprule
\small
% \hline
      & FLASH (Default) & SingleWeight &MultiOut & VEER  \\ \hline
SS-A & 0.007 & {0.011} & \cellcolor[HTML]{DADADA}{0.014} & 0.013 \\
SS-B  & 0.125 & \cellcolor[HTML]{DADADA}{0.171} & 0.100 & 0.119 \\
SS-C  & 0.020 & {0.021} & 0.019 &{0.022}  \\
SS-D  & 0.060 & {0.059} & 0.061 & 0.058 \\
SS-E & 0.131 &{0.135} & 0.144 &{0.136}\\
SS-F  & 0.098 & {0.082} & 0.113 & 0.113 \\
SS-G  & 0.014 & \cellcolor[HTML]{DADADA}{0.296} & 0.016 & 0.014 \\
SS-H  & 0.013 & {0.013} & 0.016 & 0.013 \\
SS-I  & 0.034 & {0.038} & 0.034 & 0.042 \\
SS-J  & 0.386 & \cellcolor[HTML]{DADADA}{0.437} & 0.384 & 0.362 \\
SS-K  &{0.299} & \cellcolor[HTML]{DADADA}{0.306} & 0.304 & 0.298 \\
% \hline
\bottomrule
\end{tabular}

\label{tab:rq3}
\end{table}

\begin{figure}[t!]
\centering
\includegraphics[width=.99\textwidth]{ 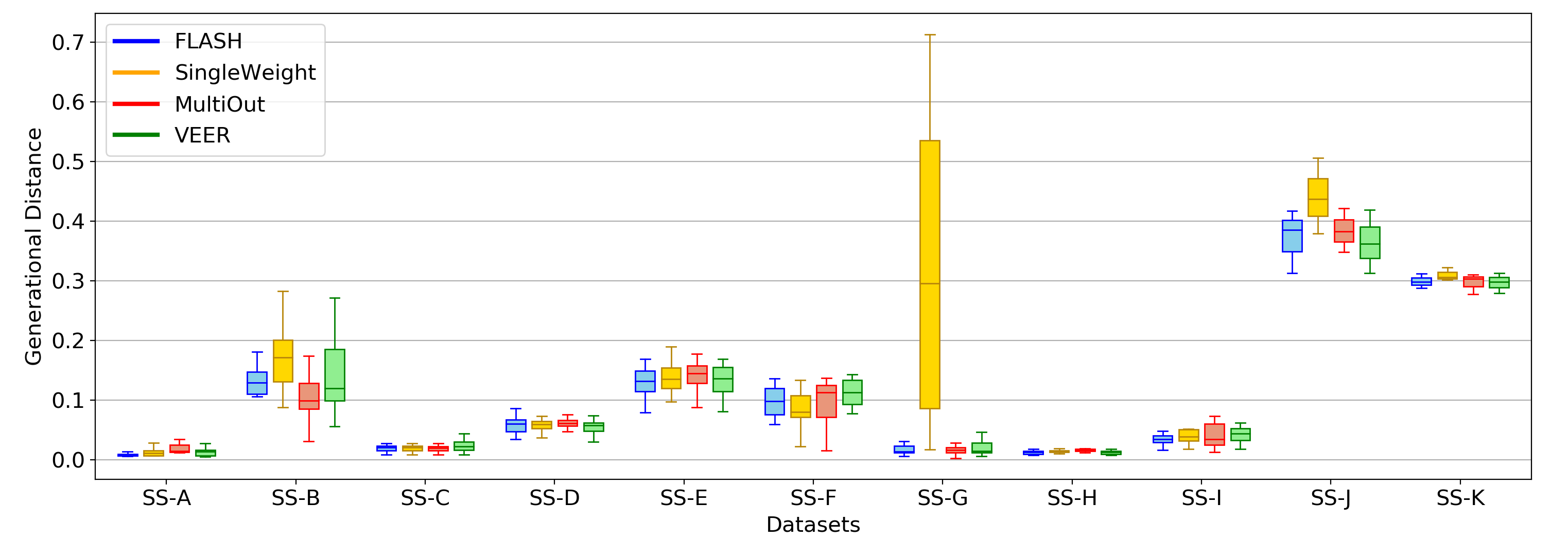}
\caption{RQ3 result: The distribution of GD measures for all four models. The GDs from the last 2 datasets are relatively higher than others, partially because these 2 datasets are (10+ times) larger than other datasets.  } 
\label{fig:rq3}
\end{figure}

\subsection{RQ3}
\textbf{Can VEER resolve the model disagreement problem while maintaining on-par performance with benchmark methods?}

First of all, we need to clarify that we will not use the \textit{rank correlation} as the major indicator to evaluate the merit of our approach. The reason is, when the model used to generate the solution set is a single-output learner, there is naturally no disagreement at all, which always guarantees a perfect correlation (Kendall's $\tau$ coefficient = 1). 
% Since the VEER model reasons about the optimal solutions (configurations) by estimating their overall performance, the configurations believed to have dominating performance in one objective are identical to those believe to have dominating performance in other objective(s). In other words, the same acquisition function is used to locate optimal (non-dominated) configurations for all the objectives. 
This result is totally to be expected by us since VEER is designed purposefully to resolve the disagreement problem.

Therefore, we need to evaluate whether VEER compromises the performance as compared to FLASH. 
In Table \ref{tab:rq3}, we report the generational distance (GD) of solution sets provided by each optimizer. We used a non-parametric effect size test to determine if the difference between the two performance measures is statistically significant.
As shown by the table, in most cases VEER can achieve comparable performance.
\BLUE
We note in Figure~\ref{fig:rq3} that VEER in some cases has a slightly larger variance than other methods. 
% That said, looking across all the results, we would summarize all these results are mostly stable (with the exception of the SingleWeight result of SS-G). 
Specifically, in four cases out of eleven (SS-B, SS-C, SS-F, SS-I), VEER has a larger inter-quartile (IQR) than FLASH. 
This was actually a pre-experimental expectation since VEER is a patch on FLASH (so VEER gets all the variance of FLASH, plus some extra "wriggle" due to its own learning process). 
It is also noteworthy that in the majority datasets (SS-A, SS-C, SS-D, SS-G, SS-H, SS-I, SS-K), the inter-quartile range of VEER was very small (less than $0.033$).
Lastly, we note that while that IQR is larger, the significance of that larger size does not effect our argument (as supported by the statistical tests in \S\ref{stats}.) 
% \respto{1a1} \respto{2a2} \respto{2a3}
\BLACK

%  That is, after combining knowledge from multiple learners into one learner, VEER has experienced no (or trivial) information loss. Additionally, as Figure \ref{fig:rq3} shows, VEER maintains the same level of robustness compared to other benchmark methods as indicated by the variance of the distributions. 

% It is also noteworthy that in some cases the naive approaches can also obtain similar performance compared to the original FLASH. Considering the interpretability of multi-objective models as reflected in Table \ref{tab:rc}, this sends us a sign that at least we can improve FLASH by using a simple variant of itself, which can achieve similar performance with a relatively low level of disagreement.
% This result also send us an important sign that even if we are handling a multi-objective task, it is still viable to transform the task into a single-objective task where the new single objective is synthesized using heuristics like the one within VEER.

In summary, we answer \textbf{RQ3} as follows:
\begin{blockquote}
% [width=\linewidth,colbacktitle=gray,title={Result 1}]  
\textbf{Answer 3}: 
     The design choices made for VEER are able to resolve the model disagreement problem. \BLUE Moreover, it does not compromise the performance of the original optimization model in most cases (in some cases, VEER obtained greater variance in GD than FLASH or other methods). \BLACK
\end{blockquote}

\begin{figure}[b!]
\centering
\includegraphics[width=.75\textwidth]{ 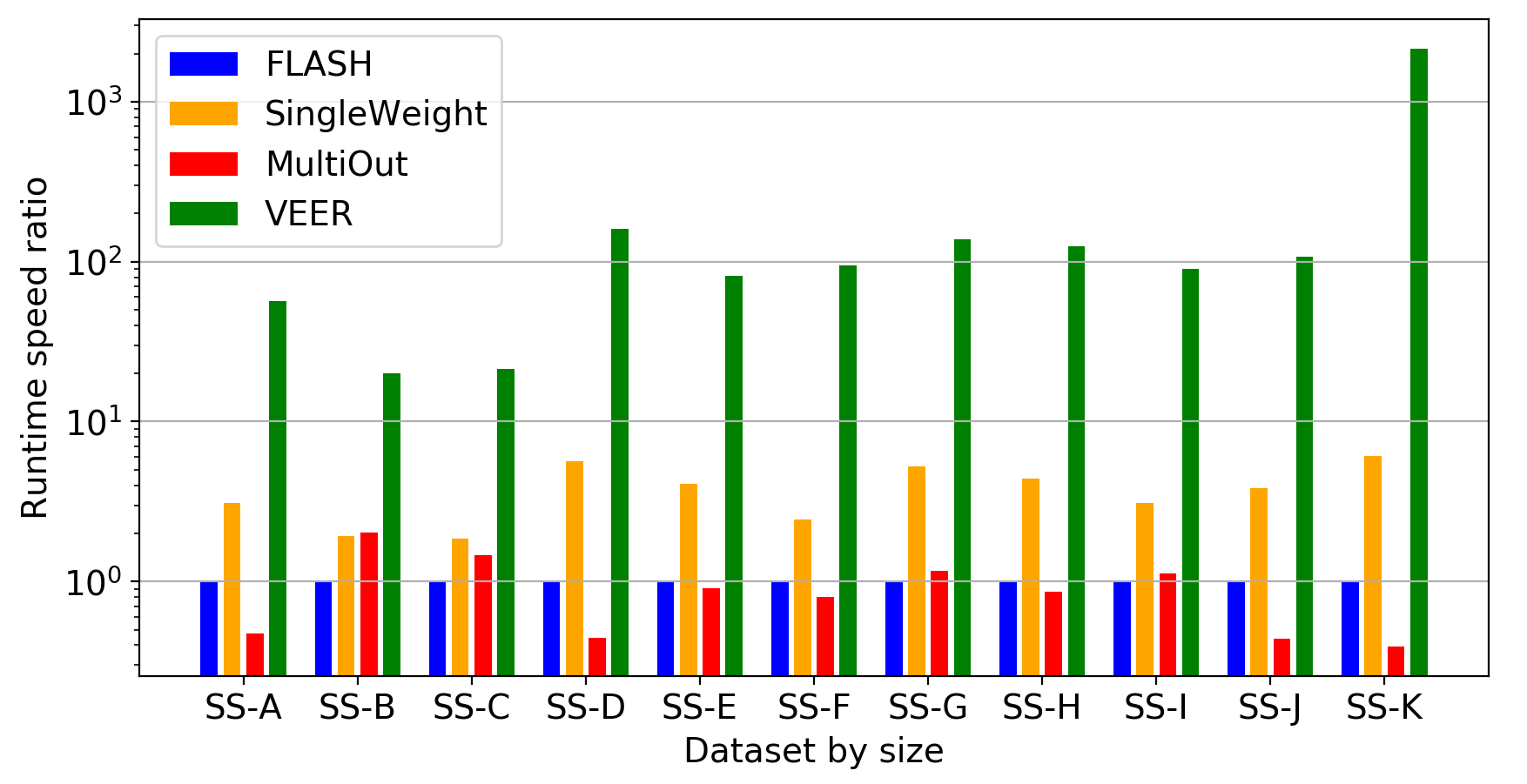}
\caption{RQ4 result: The inverse runtime ratio using FLASH's runtime as the benchmark, calculated by dividing the average runtime of FLASH over that of other methods. Higher ratio is better.} 
\label{fig:runtime}
\end{figure}

\subsection{RQ4}
\textbf{Can VEER reduce the execution time?}

While executing VEER during the experimentation, one of the most obvious bonuses is that VEER runs much faster than prior methods when applied to the holdout set. 
% This is because the algorithmic design of VEER has reducing the computational complexity of applying the optimization model. Unlike the benchmark algorithms, which generate multi-output predictions on configurations' performance, VEER only returns the predicted domination rank as the final output. This means that VEER can skip the non-domination sorting process given the output has only one dimension. 
% This is because the non-domination sorting has a runtime complexity of $O(N^2)$ where $N$ is the size of the dataset, while sorting a one-dimensional vector only takes a complexity of $O(N\log N)$.
As shown in Figure \ref{fig:runtime}, the average execution time using VEER when applied to the holdout set is much faster than that of other benchmark methods. 
The model can be up to 1000+ times faster than other benchmarks when applied on the largest dataset. Note that we also observe a larger variance in runtime when the size of datasets increases. This is because the runtime is also hugely influenced by the size of the returned solutions set: when there are more non-dominated solutions in the holdout set, the time till the termination of the non-domination sort will increase proportionally. 
% Using the largest dataset SS-K as an example, there are trials where the returned solution set from FLASH contains over 100 solutions (configurations), and the runtime of that trial will be much greater than if the solution set only contains 10 solutions.
VEER does not suffer from such complexity given its design of "compressing" multi-objective space into a single dimension.

In summary, we answer \textbf{RQ4} as follows:
\begin{blockquote}
% [width=\linewidth,colbacktitle=gray,title={Result 1}]  
\textbf{Answer 4}: 
      As shown in the experiment result, VEER takes a much shorter time to generate the solution set out of the holdout set than other methods. Moreover, as the size of the holdout set increases, the execution time of VEER grows much slower than that of other methods, which indicates better scalability of VEER.
\end{blockquote}
    
\section{Discussion}\label{discussion}
In this section we discuss what makes VEER novel and more useful than prior methods in optimizing multi-objective configuration.

{\bf Final interpretations:} In the domain of multi-objective optimization, final solutions are yielded by processing non-dominated sorting first. However, this part is non-parametric and cannot be used to extract interpretation. Therefore, traditional SMBO optimizers are only capable of providing preliminary interpretations from single-objective models. The same attribute is also reflected on Clafer, the configuration visualizer: Clafer helps practitioners understand the commonalities and variants among different configuration solutions on the Pareto frontier. However, such instance-based visualization cannot yield generalizable rules on how to optimize multiple goals.  VEER, on contrary, provides final rule-based interpretations about how to optimize a configuration in general.

\begin{figure}[b!]
\centering
\includegraphics[width=.95\textwidth]{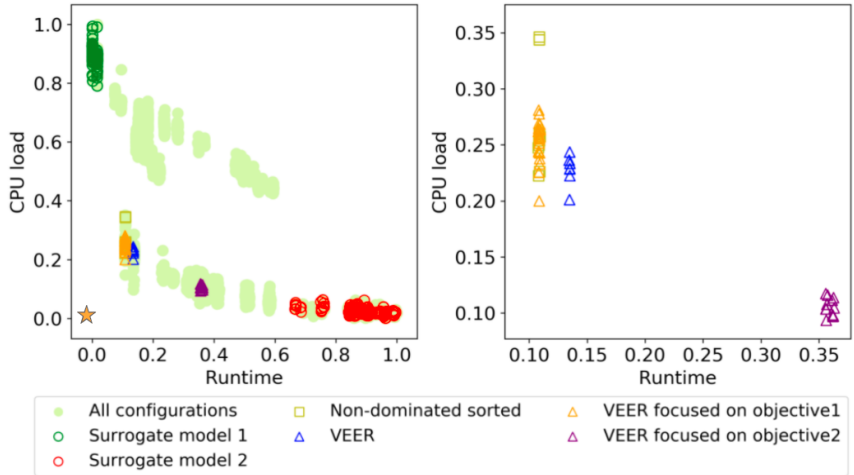}
\caption{As per Figure \ref{fig:mdp}, squares marked as ``Non-dominated sorting'' denotes the final solutions return by FLASH. Figure on the right side shows the zoom-in area of FLASH and VEER solutions. These results are better than those seen in 
Figure~\ref{fig:mdp} since it ensures interpretations from VEER are final. Moreover, VEER can be adjusted, if needed, to reflect user preferences on different objectives.} 
\label{fig:more}
\end{figure}
{\bf Fast:} Non-dominated sorting is a computation-intensive step with an optimizer. VEER replaces it with the hyper-space model that mimics the behavior of the non-dominating sorting procedure. By directly learning the relationship between the non-dominated ranks and the independent variables (configuration options), VEER can achieve significantly (up to $10^3$ in the largest system) faster execution time.

{\bf Adjustable:} In addition to the merits above, VEER can also incorporate with the preference-based approach by simply adjusting the distance calculation function insides ZIGZAG. As shown in Figure \ref{fig:more}, by customizing the preference weights (e.g., $[2:1]$ means optimizing one objective is twice as important as optimizing the other one), VEER can generate solutions that reflect such user preferences.

{\bf Uncompromising:} Our approach cannot be generalizable to all configurable software systems: when two objectives are competing (with strong negative correlation), VEER is doomed to fail since a configuration that optimizes one objective will inevitably compromise the other one by the same extent. However, we show that this is not the case in datasets explored so far. \BLUE In fact, VEER can generate disagreement-free interpretations {\em without} compromising the performance in most cases; In others, VEER inevitably suffers from greater variances.\BLACK

\section{Threats to validity}\label{threat}
Given the complexity of our experiments in 11 real-world configurable software systems, many factors can threaten the validity of our results.
\newline

{\em Internal Validity.} First of all, the multi-objective optimizer learns from benchmark measurements that we collected from various configurable systems. While we have rejected measurements with relatively high variance, it remains possible that some measurements are incorrect, which can bias the learning procedure of the optimizer and may result in worse solution sets. 

Secondly, we measure the level of model disagreement using a rank correlation test, namely Kendall's $\tau$ test. One shortcoming of this test is that it has to be performed on ranks assigned to the same set of variables, which forces us to compute the correlation on the whole population (since optimal solutions defined by different objectives are hardly the same). As is well-known, it is actually the optimal (non-dominated) solution set that an optimizer cares about, and the very majority of the whole population are only sub-optimal solutions. Therefore, it could be a possible case that there is actually little disagreement within the optimal solutions but much disagreement among the sub-optimal space. 
% Under such a scenario, Kendall' $\tau$ may inaccurately reflect the disagreement level in the area we are truly interested in.

{\em External Validity.} 
% Our results are not perfectly transferable to larger models or other configurable systems. 
While we believe our findings is generalizable as supported by the experimentation result, this does not guarantee the model to be automatically scalable to systems with larger spaces and dimensionalities. However, to increase the external validity of our study, we did intentionally choose datasets from various sizes, dimensions, and application domains.  

Secondly, we select CART as the embedded surrogate model because prior research has shown that CART is capable to achieve good performance and has great interpretability with feature interactions being presented as decision conditions within the tree~\cite{nair2018finding}. Another consideration is that since we use FLASH, one of the most recent state-of-the-art optimization models, as the benchmark method, we would prefer to control variables within the experiment so that our comparison is sufficiently "fair". That said, it is possible that other white-box machine learners (e.g., logistic regression, Naive Bayes) can achieve superior performance than CART while we have not explored yet. However, this paper is centered around assessing the feasibility of our proposed method used to improve Pareto-based optimization. Our current experiment results suffice to illustrate the effectiveness of our work.

\section{Conclusion}\label{conclusion}
We have shown that {\em model disagreement} is rampant in
the standard case studies used to assess 
multi-objective configuration problems.
As stated in our introduction,
the current literature
has surprisingly little on this topic.
Hence we are concerned
that 
model disagreement may be a   long-standing, but previously under-explored, problem.

To better address this problem,
we have proposed a confusion-free multi-objective configuration optimizer, VEER, which is built on top of a state-of-the-art sequential model-based optimizer FLASH. We have shown that VEER has not only inherited many merits of FLASH (good performance and low training cost), but also resolved the potential model disagreement problem. We have demonstrated the effectiveness of VEER in resolving model disagreement while maintaining on-par quality for the configuration solutions.

To do this,
first, we investigated the existence of model disagreement problem in cases studied in this paper, where the interpretations returned by FLASH can sometimes be conflicting, as indicated by the rank correlation. 

Second, we demonstrated that VEER can enhance the shortcoming of FLASH. VEER is capable of mapping an N-dimensional objective space (in this paper, $N$ is 2 and 3) onto a single-dimensional space without information loss of the domination relationship. Since the synthetic single-dimensional objective space can be learned by just one machine learner, the model disagreement problem collapses in itself. 

Finally, we have  shown that VEER can achieve on-par performance compared to the original FLASH model, indicating no information loss during the procedure. 

Another bonus of VEER is that by simplifying the objective space, the execution time of applying the optimizer model during the testing (or deployment) time has been dramatically reduced (1,000 times at most). It is also noteworthy that since the computational cost of VEER is also relatively small compared to that of an optimizer, the overhead of adding VEER on top of any model-based optimizer should be trivial.  

\BLUE
Regarding future work, we will invest our effort in addressing open issues described in \S\ref{threat}. Beyond that, we believe VEER can be not only applicable in configuration optimization tasks, but also hyper-parameter tasks from machine learning techniques as described in the FLASH paper~\cite{nair2018finding}. Moreover, there exists an increasing demand for commissions such as auto-generated code (from chatGPT or Copilot, etc.) to the local domain. Tools like VEER might be useful for tuning the choice points inside that code (e.g., given a genetic $k$-th nearest neighbor code snippet from chatGPT, designers could use VEER to decide what is the best setting for $k$.
% \respto{1a2} \BLACK

% To assess interpretability in multi-objective tasks is a relatively under-explored practice area, so our primary focus is to build more baseline methods to defined and evaluate interpretability in a more comprehensive manner. Specifically for our approach, more optimization work needs to be done to further test the robustness of VEER: We should verify if better hyper-parameter tuning can significantly improve the performance of VEER. We also need to check if VEER is applicable on machine learners other than CART. Moreover, we would like to test VEER on larger systems with more configurable options. 

\section*{Acknowledgements}
This work was partially funded by 
a research grant from the Laboratory for Analytical Sciences, North Carolina State University.
Apel's work has been funded by the German Research Foundation (AP
206/11 and Grant 389792660 as part of TRR 248 – CPEC.
Siegmunds work has been supported by the Federal Ministry of Education and Research of Germany and by the S\"achsische Staatsministerium f\"ur Wissenschaft Kultur und Tourismus in the program Center of Excellence for AI-research "Center for Scalable Data Analytics and Artificial Intelligence Dresden/Leipzig", project identification number: ScaDS.AI, and by the German Research Foundation (SI 2171/2-2).

\section*{Declarations}
\subsection*{Funding and Conflicts of interests}
Apart from the funding acknowledged above, this work does not have any other conflicts of interests.

\bibliographystyle{spbasic} 
\bibliography{main}

\end{document}